\documentclass[conference]{IEEEtran}
\IEEEoverridecommandlockouts
\usepackage{cite}
\usepackage{amsmath,amssymb,amsfonts}
\usepackage{algorithmic}
\usepackage{graphicx}
\usepackage[shortlabels]{enumitem}
\usepackage{textcomp}
\usepackage{xcolor}
\usepackage{url}
\usepackage{multicol}
\usepackage{cleveref}
\usepackage{stfloats}

\usepackage[acronym]{glossaries}
\newacronym{ai}{AI}{Artificial Intelligence}
\newacronym{ml}{ML}{Machine Learning}
\newacronym{dl}{DL}{Deep Learning}
\newacronym{lts}{LTS}{long-term support}
\newacronym{gpu}{GPU}{Graphics Processing Unit}
\newacronym{capex}{CapEx}{Capital Expenditures}
\newacronym{opex}{OpEx}{Operation Expenditures}
\newacronym{OI}{O\&I}{Operations and Infrastructure}

\def\BibTeX{{\rm B\kern-.05em{\sc i\kern-.025em b}\kern-.08em
    T\kern-.1667em\lower.7ex\hbox{E}\kern-.125emX}}
\begin{document}

\title{\textbf{\textit{Beyond Desktop Computation:\\Challenges in Scaling a GPU Infrastructure
	}} \huge \\
\thanks{\hrule height 0.5pt \vspace{4pt} Martin Uray and Eduard Hirsch were supported
	by the European Interreg Österreich-Bayern project AB215 DataKMU and Michael
	Gadermayr was supported by the County of Salzburg under grant number FHS-2019-10-KIAMed.}
}

\author{
\IEEEauthorblockN{Martin Uray}
\IEEEauthorblockA{Salzburg University of Applied Sciences\\Salzburg, Austria\\
martin.uray@fh-salzburg.ac.at}\\   
\IEEEauthorblockN{Gerold Katzinger}
\IEEEauthorblockA{NTS Netzwerk Telekom Service AG\\Salzburg, Austria\\
gerold.katzinger@nts.eu}
\and
\IEEEauthorblockN{Eduard Hirsch}
\IEEEauthorblockA{Salzburg University of Applied Sciences\\Salzburg, Austria\\
eduard.hirsch@fh-salzburg.ac.at} \\
\IEEEauthorblockN{Michael Gadermayr}
\IEEEauthorblockA{Salzburg University of Applied Sciences\\Salzburg, Austria\\
michael.gadermayr@fh-salzburg.ac.at}

}

\maketitle

\begin{abstract}
Enterprises and labs performing computationally expensive data science
applications sooner or later face the problem of scale but unconnected
infrastructure. For this up-scaling process, an IT service provider can be hired
or in-house personnel can attempt to implement a software stack. The first
option can be quite expensive if it is just about connecting several machines.
For the latter option often experience is missing with the data science staff in
order to navigate through the software jungle. In this technical report, we
illustrate the decision process towards an on-premises infrastructure, our
implemented system architecture, and the transformation of the software stack
towards a scaleable \gls{gpu} cluster system.
\end{abstract}


\section{Introduction}
In the course of computing history, sufficient computing power often exhibited
a basis for the application of novel ideas.
But it often also has been the basis for research ideas that were not superior
because of the idea, but simply because of sufficient computing power available.
Here, in some kind of sense, the increasing computational power drove the
development of research ideas\cite{hooker_hardware_2020}. The steadily increasing
amount of computational power was long driven by Moore's Law, whereas the
producing industry is not driven by this observation
anymore\cite{waldrop_chips_2016}.
Also the field of \gls{ai} gained  popularity due
to the amount of increased computational power, and evolved in new fields, like
\gls{dl} \cite{simonyan_very_2015,krizhevsky_imagenet_2017}.

Due to the promising results by applications of \gls{ai} methods, a lot of research
is currently performed in the field of \gls{ai} based methods. Working with \gls{ai},
especially \gls{dl}, the question of the computational resources arises sooner or
later. We observed, that in a lot of companies, Data Scientists are often the
only ones with expertise in Computer Science. Besides their main function, they
have to take care about computing infrastructure - even if it is not their field of
expertise. This causes, that a huge amount of time is spent on research for
applicable tools, systems, and environments to develop on.

Similar to other companies, start-ups, and universities, a simple \gls{gpu}
on-premises infrastructure is maintained within our institute, consisting of
only a single server. 
This computing infrastructure was initially implemented as simple as possible,
due to the lack of expertise and time. By using fair-share, one server with eight
\gls{gpu}s was set up. This machine specification can be found in
\Cref{tab:specs}, line $C_1$. On this machine, for eligible people access was
granted for research and training. When submitting a computation job, one user
had to choose a not used \gls{gpu} and start the job by only using this one
dedicated resource. For multi-\gls{gpu} jobs, several resources were allocated
and blocked. With this setup it was likely to happen, that one user did not
restrict a jobs resources according to the policy and blocked all computational
resources. Additionally, inferences with other jobs are possible, leading to
failures in all involved jobs. With a growing number of users on such a shared
hardware, the demand for manageable permissions and restrictions increased.

Up-scaling and extending such an on-premises infrastructure, while preserving
easy usage can be quite complex, without the help of experts in the field of
IT infrastructure and High-Performance Computing. Also, the decision, on
whether to move to the cloud shall be well defined, since this must be argued
towards various stakeholders.

In the progress of planning the extension of our existing computational resources,
several questions came up while being aligned with the set requirements:

\begin{enumerate}[label=Q\arabic*]
	\item Is a transition into the cloud financially beneficial or should the
		  existing on-premises infrastructure be extended?
	\item Are there preexisting solutions available?
	\item What hardware components are needed and how to design the architecture?
	\item What software components need to be part within the used software stack?
\end{enumerate}

In this work, we are outlining the architecture of our established
\gls{gpu} computing infrastructure, as it scaled from a single server to a
multi-instance computing cluster. This is based on the decision towards an
on-premises infrastructure and the alignment with defined requirements. This
report shall not give the impression of being a best practice, however, it is
intended to show the considerations that are necessary when transforming
infrastructure to a multi-instance cluster setup.

This report is structured as follows: In \Cref{sec:cloud-vs-onpremises} we give
a short overview of our considerations to decide on a Cloud or on-premises
computing infrastructure within our institution. \Cref{sec:req} outlines the
requirements on the infrastructure, that led to the architecture of the
infrastructure \Cref{sec:infra-arch} and the setup of the cluster
\Cref{sec:cluster-setup}. In \Cref{sec:discussion} we discuss about advantages
and disadvantages of our implementation with an outlook and further
considerations.

\section{Cloud vs. On-premises Computing}
\label{sec:cloud-vs-onpremises}
Deciding on weather to run ones projects in the cloud or as an on-premise
infrastructure involves many different aspects which are examined in the
following. As one question this report shall answer is weather a transition
for the computational tasks into the cloud is beneficiary, the emphasis in
this section is put on factors which could lead or force ones intention to one
or the other solution.

Like described in \cite{baschab2007executive}, there are models for grasping
the scope of IT activities which shall not be our main focus here. Nevertheless,
finding an appropriate solution for the own institution or company is key to a
cost-effective solution. When further referring to \cite{baschab2007executive}
not just IT based \gls{OI} need to be considered but also ones that influence those.
Thus, each of these may change when adopting from one model to another. Due to
these manifold influencing factors in those areas and activities, a decision is
highly based on the processes an institution applies.

Some of the important key-factors are: Cost, Scalability / Upgradeability, Network
Connectivity, Maintainability, Security, General Data Protection Regulation,
Disaster Assistance, and Data Backup and Recovery. These factors are essential
to consider when planning to deploy services to cloud providers. Nevertheless,
\textit{cost} is often the one considered first.

\subsection{Costs}
\label{ssec:costs}

When comparing Cloud to On-Premise expenses, companies usually start looking at
hardware bought for local usage and on-demand (virtual) hardware. But it is
necessary to get a complete picture of the total costs which cannot be reduced
to a plain procurement process.

As indicated in \cite{baschab2007executive}, \gls{OI}
needs to be accounted for which translate to \gls{capex} and \gls{opex} when
referring to economical terms. Additionally, it is noted, that especially for
maintenance, it is necessary to spend a regular, potentially high amount for
an on-premise solution compared to a similar cloud resource. This is based on
the companies needs: electricity costs, administrative staff, licenses
and trainings.

Cloud resources, therefore, are incorporating those additional costs, offering
services at a fixed price per resource and consumed time. However, it is still
possible, to create solid cost efficient structures for machine learning when
reducing \gls{opex} and keeping \gls{capex} at an acceptable level. Especially
when dealing with \textit{Green AI}. That may be the usage of renewable energy
sources or facilitating outdated refurbished machines/hardware, in order to
establish a \gls{gpu} cluster. 

For the decision on the extension of the institutes infrastructure experiments,
regarding the usage of the existing infrastructure, where conducted. Over the
period of four months (April - July 2021) the usage was monitored and logged.
The usage over this period is used as a baseline\footnote{Pessimistic baseline,
    since during the second half of the year (Aug - Dec) a much higher usage is
    observed.} for the calculation process. Overall $7366$ hours of GPU usage
were monitored, what translates to a $32\%$ grade of operation.

\begin{table}[]
\centering
\caption{Rough estimate in EUR when On-Premise solutions pay off.\protect\footnotemark}
\label{tab:cost-overview}
\begin{tabular}{llllll}
\textbf{Month}           	&	 \textbf{1} 	&	 \textbf{2} 	&	 \textbf{3} 	&	 \textbf{11}     	&	 \textbf{16}	 \\ \hline
Procurement              	&	24500	&		&		&		&		\\
Electricity              	&	250	&	250	&	250	&	250	&	250	\\
Manpower                 	&	1200	&	1200	&	30	&	30	&	30	\\
Cost p. m. On-Premise    	&	25950	&	1450	&	280	&	280	&	280	\\
Overall On-Premise       	&	25950	&	27400	&	27680	& \textbf{29920} & \textbf{31320}  \\ \hline
Cost p. m. Google        	&	2057	&	2057	&	2057	&	2057	&	2057	\\
Overall Google           	&	2057	&	4114	&	6171	&	22627	& \textbf{32912}  \\ \hline
Cost p. m. Azure         	&	2947	&	2947	&	2947	&	2947	&	2947	\\
Overall Azure            	&	2947	&	5895	&	8842	& \textbf{32420} &	47156	\\

\end{tabular}

\end{table}

\footnotetext{Rounded to full numbers. Break-even points indicated in bold.}

In \Cref{tab:cost-overview} an estimation of the costs over a period of $16$
months is shown. The procurement costs are based on the existing on-premise
infrastructure ($C_1$). The costs of electricity are estimated based on the
assumption, that $60\%$ of the maximum power consumption is used in idle mode
and the rest added based on the actual usage.\footnote{Not covered in           
    calculations: $5\%$ of the electricity consumed in-house are produced
    with solar panels.}
The assumption on manpower is based on the fact, that the initial setup was done
during the first two months and further maintenance was barely needed. With this
on-premise solution, the initial costs were high, but the overall monthly costs
were kept low.\footnote{The hourly rate for manpower has been set
    to 30€/h. The server is running at 2600 Watt and the electricity prices set
    at 0,28€/kWh.}

For the calculation of the monthly costs, the mean monthly usage is considered
($1843$h/month). For both, MS Azure and Google Cloud, a setup was chosen with
a commitment to min. three years.\footnote{
    Prices (in EUR) as of 05.08.2021. Azure configuration: \textit{NC24s v3}
    instances with 24vCPU, 448 GiB Ram, 4x Tesla V100. Google Cloud:
    \textit{AI platform} with \textit{BASIC\_GPU} training tier. All platforms
    hosted in Europe.}

Comparing all three variants (on-premises, Google Cloud and Microsoft Azure)
with the observed rate of operation, on the "costs per month" the Azure
configuration has a break-even at month 11 and the one from Google Cloud
from month 16. Although, these systems are not completely equal in terms of
hardware, the table provides a rough direction when an on-premise solution
gets profitable.

\begin{figure}
    \centering
    \includegraphics[width=\columnwidth]{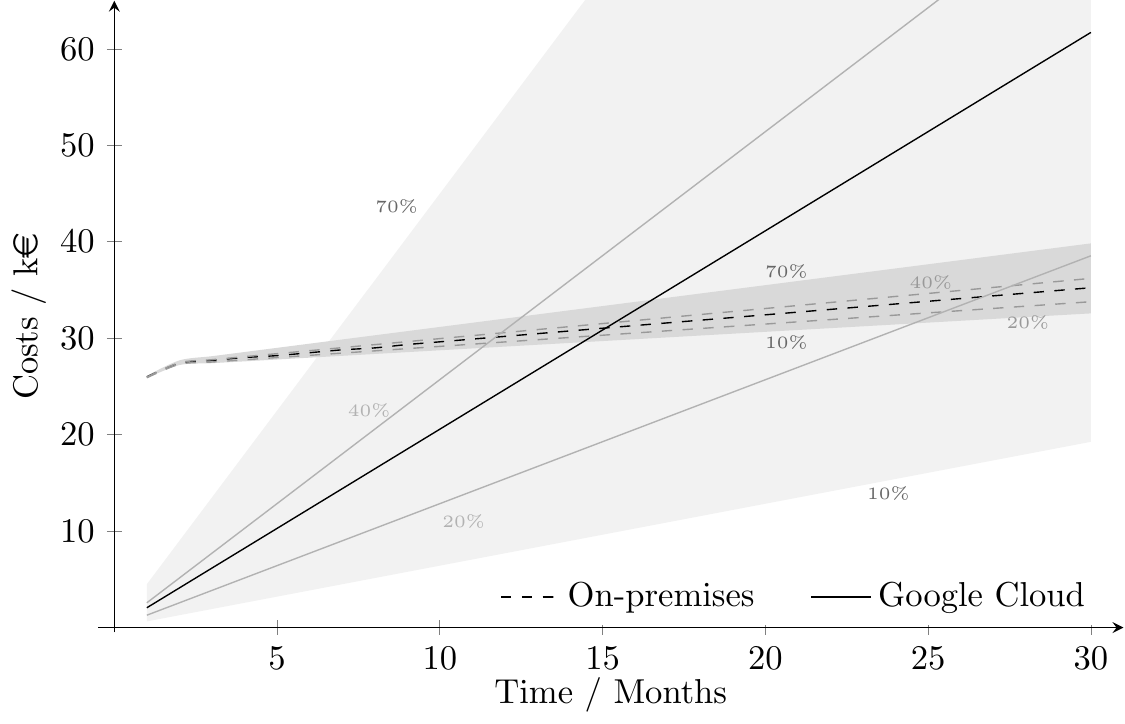}
    \caption{Cost tendency graph, illustrating the costs per usage for the used
        on-premises (dashed line) and a fictional cloud solution (solid
        line) for the calculated usage of the cluster. The shadowed areas
        denote varying usage in the range of $10-70\%$. For each,
        on-premises and Google Cloud, a rough estimation for the usage of
        $20\%$ and $40\%$ is indicated.}
    \label{fig:costtrend}
\end{figure}

\Cref{fig:costtrend} illustrates the same data, where the black solid and dashed
line represent the costs for the Google Cloud and on-premises solution,
respectively, over time. The shadowed area around both lines indicate a
calculated range of variance for a mean usage between $10\%$ and $70\%$.
Additionally, for both variants two further lines, indicating the theoretical
trend for $20\%$ and $40\%$, are indicated by the gray lines.

More specifically, \Cref{fig:costtrend} renders the trade-off between cloud and on-premises
computing in terms of costs. The on-premises solution faces the issue of high
costs in the initial phase, but far less further costs. Using cloud services
one pays exactly what is consumed. Depending on the actual usage, on-premises
can pay-off sooner or later in case the system set-up may allow to save expenses
during procurement or the running ongoing costs.

In literature also other comparisons on the costs of cloud vs. on-premises
computing can be found, see \cite{fisher2018cloud}.

\subsection{GDPR}
\label{ssec:protect-data}

The protection of the intellectual properties and values of an institution or conforming to GDPR can be one of the reasons which force professionals to refrain from using cloud resources. Although, the large established cloud providers (e.g. Microsoft, IBM, Amazon, Google) provide possibilities to rent services  which are hosted in particular regions, it is important to note that those are mainly American companies which may be forced to provide information, possibly sensitive data, even if stored in some other a non-us region.

\subsection{Other issues with Cloud Resources}
\label{ssec:other-issues-cloud}

In the following other important issues when dealing with cloud resources are listed: 

\begin{itemize}
    \item Attacks: cloud resources might be more vulnerable to attacks, not because they are more vulnerable in principle but with services, domains and exploits well-known by hackers they are rather under attack than custom services unknown to the (international) public.
    \item Network connectivity dependency and downtimes: Challenges arising due to technical outages of the world wide web may severely brake your manufacturing e.g. when thinking about production processes demanding real-time processing.
    \item Limited control and flexibility: Although, cloud providers getting richer in terms of tools and services, controlling remote hardware, servers or services may be still quite hard to deal with and may come with their own peculiarities which are difficult to manage.
    \item Vendor lock-in: provided services and interfaces are often highly proprietary and do not conform to cloud native structures or even basic IT standards.
    \item Disaster Assistance and Data Backup and Recovery may grow in complexity if e.g. a non-cloud based backup is required for cloud resources. However, for the GPU cluster backup only exists in form of a recovery image which restores the initial installation state. Further, as this cluster is used as a computing engine, which is doing batch processing, data on that cluster also resides on other systems and is not lost when data corruption occurs.
\end{itemize}

\section{Requirements to our on-premises infrastructure}
\label{sec:req}
The intended infrastructure is used for different purposes (research, course work) and by different groups of people (faculty, scientific staff, and students). Each of the stakeholders has different requirements for the setup. Before the design of the architecture, the requirements were defined in order to select the components of the software stack and the topology of the cluster accordingly. For our setup, we identified requirements, which are discussed in the following.

\begin{enumerate}[(A)]
    \item \textbf{Ease of usage}
    \label{ssec:ease-of-usage}
    Development of all algorithms and jobs to be executed is done on local machines.
    To execute the jobs on the cluster, no modifications on code and no major
    changes on the call are necessary. Data and other necessary
    resources need to be available for the job, where ever it is scheduled to be
    executed. Additionally, no knowledge about the system itself is necessary for
    execution.

    \item \textbf{Scheduling of Jobs}
    \label{ssec:schedule}
    A jobs execution is decoupled from the scheduling. This means, that
    a users, who submits a job, does not have to know about the architecture of
    the cluster or the setup of the computing nodes.
    When submitting a task, this is appended to a queue of jobs. When
    executing a job, the request for a certain type of resource is stated (e.g.
    type of \gls{gpu}, memory, etc.). The system schedules the jobs according to
    first-in-first-out and the availability of the requested resources.

    \item \textbf{Workload Distribution}
    \label{ssec:workload}
    Resources shall be used equally over the system, where possible. If several
    computing nodes with likewise configuration exist, it must be assured that
    the load of work is distributed among those nodes.
    It must be avoided, that only few components take care of the whole computation,
    whereas the others are not used.

    \item \textbf{Permission Management}
    \label{ssec:permissions}
    Different stakeholders use the cluster for different purposes. Hence it shall be
    possible to assign users to groups to restrict their usage to a defined policy.
    As a figurative example, members of the group \textit{students} shall only be
    allowed to use $x$ \gls{gpu} at a time for a maximum of $n$ hours. This shall
    reduce the potential risk of misusing the infrastructure.

    \item \textbf{Maintainability and Scalability}
    \label{ssec:maintain}
    It shall be possible to remove and exchange parts of the system, without too much
    time effort, and downtime. The same applies to the extension of the system: For
    future developments, the system must be designed in such a way, that additional
    computing power can easily be extended with no major modification to the system
    itself, which enhances complexity to the architecture and the configuration.
    Heterogeneous setups should be configurable.

    \item \textbf{Network Speed}
    \label{ssec:speed}
    When designing the topology of the connection between all the components within
    the system, it shall be taken care of to design this system in such a way, that
    the network speed is high and influences from other network traffic that reduce
    speed kept low. On the other hand, also the traffic between the components must
    not have any influence on the other resources on the network. Printing, access
    to the web or file servers must not be reduced in speed, just because of
    transferring data among two nodes.

    \item \textbf{Costs}
    \label{ssec:req_costs}
    The last requirement concerns the monthly costs of the infrastructure. Monthly
    costs are intended to be as low as possible, while initial costs
    must be below a certain predefined budget.

\end{enumerate}

\section{Infrastructure Architecture}
\label{sec:infra-arch}
During the research, several solutions, offered by various companies were
found, like \cite{amette_scalable_2019}. All of these solutions posed the
drawback to exceed the defined procurement budget.

\begin{figure}
    \centering
	\includegraphics[width=\linewidth]{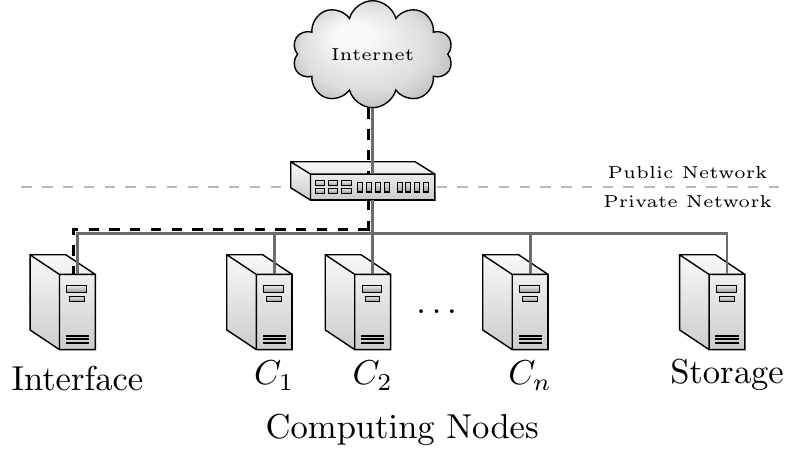}
	\caption{The network architecture of the setup contains an interface node,
		several computing nodes ($C_1 \ldots C_n$), and a storage node,
		all hidden behind a router. The dashed lined indicates the default
		route for the ssh connection.}
	\label{fig:arch}
\end{figure}

The architecture of the up-scaled infrastructure is motivated by
\cite{gupta_how_2013} and inspired by \cite{amette_scalable_2019}. The designed
architecture is illustrated in \cref{fig:arch}. The whole cluster is completely
abstracted from the public network by an interface router. This interface router
creates a private network for the whole infrastructure, without putting load on
the public network by data synchronization between the nodes. As a result, this
hides the complexity of the infrastructure to the outside, while being still
accessible from the public network. For the implemented cluster, a
\textit{MikroTik CRS-309-1G} is being used.

For the private, cluster network, instead of a conventional \textit{Gigabit
Ethernet}, a \textit{SFP+} connection is established. This protocol support data
rates up to $10$ Gbit/s. This enables faster synchronization between the nodes.

Within this private network, several nodes are connected. The interface node
poses as the entry point to the cluster and the only accessible machine within
the network for users. On this device, all tasks are scheduled and distributed
according to the set policy and configuration. This node does not have any
\gls{gpu}, so no accidental blocking of resources may happen. For the setup,
the interface node is the standard gateway for the \textit{SSH} access.

The actual computing tasks are scheduled to the computing nodes ($C_1 \ldots
C_n$). These machines are equipped with sufficient computing power, \gls{gpu}s,
and memory. For the implementation, the already available machine (old
infrastructure) $C_1$ is being used. Additionally, a further machine $C_2$ is
added to the system. The configuration of all machines ($C_1$, $C_2$, and $I$)
are described in \Cref{tab:specs}.

As indicated, this architecture also makes it easy to increase the number of
further nodes. In \cref{fig:arch}, a dedicated storage node is
indicated. This storage is available by all components within the cluster and
enables access to all data by all nodes. For the current implementation, this
storage is not included in the setup. A dedicated storage will be added with one
of the future extensions. Currently, only the storage from the nodes $C_1$ and
$C_2$ is used by all nodes.

\begin{table*}
	\centering
    \label{tab:specs}
    \caption{Configuration of the nodes as implemented within the setup.}
	\begin{tabular}{c|cccc}
		\textbf{Node}      & \multicolumn{1}{c}{\textbf{\gls{gpu}s}} & \multicolumn{1}{c}{\textbf{CPUs}}       & \multicolumn{1}{c}{\textbf{RAM}} & \multicolumn{1}{c}{\textbf{Local Storage}} \\ \hline
		I                  & -                                       & 1x AMD EPYC $7302P$, $3.00$ GHz         & $126$ GB                         & $460$ GB system                            \\
		C\textsubscript{1} & 8 NVIDIA GeForce RTX 2080Ti             & 2x Intel Xeon Silver $4114$, $2.20$ GHz & $230$ GB                         & $230$ GB system + $8$ TB data              \\
		C\textsubscript{2} & 3 NVIDIA RTX A6000                      & 2x AMD EPYC $7452$, $2.35$GHz           & $512$GB                          & $240$ GB system + $8$ TB data              \\
	\end{tabular}

\end{table*}

All devices and nodes in the implemented setup are chosen, so that they are
easily built into a existing server rack within the in-house server room.

\section{Cluster Setup}
\label{sec:cluster-setup}
For the setup of the cluster and the software stack, extensive research was
needed. This was caused by the variety of software products available and their
interoperability, enterprise products on the market, and expensive reference
solutions. The following setup is inspired by the huge computing infrastructure
of the University of Massachusetts \cite{noauthor_gypsum_nodate}. The selection
and installation of the software stack is based on the instructions in
\cite{george_slurm_gpu_ubuntu_2020}.

Each node in the initial setup of the cluster is based on a \gls{lts} Linux
distribution. In the setup a \textit{Ubuntu 20.04 LTS Server} is used. The
decision towards a \gls{lts} version was done for a long lifetime support.
Among others, consumer options used in practice are
\textit{CentOS Linux 7} \cite{noauthor_gypsum_nodate}, and \textit{Ubuntu 20.04
LTS} \cite{george_slurm_gpu_ubuntu_2020}. A comprehensive overview and
comparison of the operating systems and components used in other clusters
can be found on the \textit{TOP 500}\footnote{\url{https://www.top500.org/}} list.

On each of the computing nodes, the appropriate CUDA driver is installed.
The correct driver and CUDA version is dependent on the \gls{gpu}
and operating system.

Since the jobs to be executed are scheduled by a central entity on a machine
within the cluster, storage synchronization is crucial. A user is only able to
access the interface node. All data and environments also need to be synchronized
to the other nodes, so that the execution of the scheduled jobs work properly.
As noted in \Cref{sec:infra-arch}, no dedicated storage node is implemented within
the cluster. The storage integrated in both computing nodes, $C_1$ and $C_2$, is
provided as storage for the cluster. The storage is used as a ZFS (Zetta
file system) and mounted from all other nodes within the cluster. Each machine
mounts that storage to the same mount point in the file system. With other setups
different approaches can be found, where either everything is mounted using only
one directory \cite{george_slurm_gpu_ubuntu_2020} or the storage split among
several directories for different purposes (e.g. home directories, research
storage, scratch space, temporary Space)\cite{noauthor_description_2021}.
Within the implemented cluster, only the home directories and a data directory are
shared within the cluster.

For password-less connections from the master to all worker nodes, the ssh keys are
exchanged and munge is used as an authentication service. For the application of
the used workload manager, MariaDB is used.

As a workload manager, the decision was made towards SLURM \cite{goos_slurm_2003}.
Another workload managers commonly used is HTCondor \cite{noauthor_computing_2021}.
The decision towards SLURM is not only based on the reference implementations, but
also since SLURM is easy to use. Let's consider a simple script, name
\verb|execute_me.py|. Using SLURM, applying a standard configuration, it is
scheduled as simple as \verb|sbatch python execute_me.py|. No further adaptations
are needed to be taken care of.

Additionally, SLURM, which has a large community, is well documented, and even has
enterprise support. Furthermore, over $60\%$ of all supercomputers use SLURM in
their setup \cite{padua_top500_2011}. SLURM can be used for any size of clusters and
works well with over $1,200$, partially different, \gls{gpu}s
\cite{noauthor_gypsum_nodate} and to smaller setups and clusters.
Both, HTCondor and SLURM, can be used interoperable. An overview on this and on both
tools itself is given by \cite{hollowell_mixing_2017} and \cite{du_feasibility_2019}.

SLURM is a highly configurable component. This workload manager can be configured
in the most basic approach, where it works only as a simple workload manager that
balances jobs on all nodes, up to a system were it is a part of a software stack,
that not only restricts access to resources for certain users but also
interconnects with other plugins like for accounting for the actual usage. For the
implemented system, the initial configuration has two groups of users, where
different types of \gls{gpu}s are restricted to certain user groups. For instance,
users of the group \textit{factuly} may use all \gls{gpu}, and members of
\textit{students} are restricted to \gls{gpu}s on computing node $C_1$.
Additionally, users from the first group can schedule as much jobs as needed, where
the jobs are executed as soon as possible, and members of the latter can only have
one job running at a time.

Users and groups are managed using
\textit{FreeIPA}\footnote{\url{https://www.freeipa.org}}. This software component
offers an intuitive Web-UI for managing all accounts. For the initial setup, the
clusters user accounts are intentionally independent of the organizational
accounts for the rest of the official IT infrastructure. \textit{FreeIPA} allows
not only to add and delete users and groups, but it also allows to set validity
periods, storage quotas, and much more.

\section{Future Adaptions}
\label{sec:future}
The purpose of this cluster is in computational power for research and academia.
In both fields, it will be likely that the demand for computational power will
increase in the following years. For this purpose, several adaptions to the cluster
are planned already.

The first extensions on the number of computation nodes will be with a more
`low-spec' hardware. Using already available consumer hardware, the cluster will
be extended, so that the number of computational devices increases. Using, e.g.
several discarded, slower hardware can be offered in exchange
for more jobs in parallel.
Given the current situation on the international market, where only a small number
and on overpriced \gls{gpu}s are available, this is a very attractive option as
long as the memory requirements of jobs are not too high.
Additionally, from an economical perspective, this gives the hardware a second
life instead of being recycled.

Depending on future usage, also more memory computational tasks may be executed
on the machine. Therefore, SLURM also needs to be configured, so that no GPU, but
a certain amount of memory can be allocated.

As noted in \Cref{sec:cluster-setup}, no dedicated storage is used within the
implemented cluster. A further extension, where dedicated storage is aimed. This
storage is also planned to be fault-tolerant, by using formats of storage
virtualization.

Maintaining a continuously growing infrastructure, it is not easy to keep an
overview of the status on all machines and devices in the cluster simultaneously.
To make administration easier, a cluster management system will be implemented.
Observing the status of all connected nodes and devices, managing them,
and taking actions on events will be the main task of the component
implemented by this extension.

In \Cref{sec:cloud-vs-onpremises}, the comparison of the costs of using the
cloud infrastructure is elaborated. Depending on the usage of the cluster, it
is a further use case to overcome performance peaks by including cloud
resources by an external vendor/provider. By starting a cloud instance while
the cluster is in high demand, quick jobs can be outsourced to the cloud, which
speeds up the execution of jobs.

Depended on future usage, it is also possible to improve the hardware to enable
computational jobs, with a memory demand higher than the memory of a machine.
RDMA\footnote{Remote Direct Memory Access} is an extension technology with the
ability to access memory from one host to another remote.

\section{Discussion and Conclusion}
\label{sec:discussion}

In this work, we focused on an overview of the transformation of a research
machine with several \gls{gpu}s to an extensible cluster of several computing
nodes, each offering several \gls{gpu}s.

The requirement on the ease of usage (requirement (req.) \ref{ssec:ease-of-usage}) is preserved. Due
to the usage of SLURM, no modifications need to be done on the code or the call
itself, only the execution needs to be done using SLURM. Also, the scheduling of
jobs (req. \ref{ssec:schedule}) and the workload distribution (req. \ref{ssec:workload}) are
handled by SLURM. The permission management (req. \ref{ssec:permissions}) is also
covered, commonly by SLURM and \textit{FreeIPA}.

The requirements on maintainability and scalability (req. \ref{ssec:maintain}), as well
as on network speed (req. \ref{ssec:speed}) are both covered by the design of the
clusters architecture.
Further machines can be easily added with a minimum configuration effort. So
even when starting with a small number of computing nodes, like in our case with
two, it exhibits an effective and efficient platform to work with. The maximum
speed within the network is above the standard network used within our institution
and the traffic on the cluster does not influence the public network, due to its
private scope.

The costs (req. \ref{ssec:req_costs}) are kept low and within budget. One might argue,
that the components in this proposed setup are still not from a low-cost price
range. However, all components described within this work can be replaced with
existing and more budget-friendly components. Additionally, the monthly costs
are controlled, without surprises by an unexpected increased demand.

In the introduction, we posed four questions that were asked before this project
and answered within this work. Q1 was concerned, whether it is beneficial
for the proposed use case to use cloud resources. In
\Cref{sec:cloud-vs-onpremises}, this topic is elaborated extensively. By showing
a comparison of the costs, the decision was made against cloud solutions and to
foster an on-premises infrastructure. Also, preexisting solutions (Q2) were not
considered, since our research showed, that those solutions are above the given
budgets limit. The design of the architecture (Q3) and the components of the
software stack (Q4) are shown in \Cref{sec:infra-arch} and
\Cref{sec:cluster-setup}, respectively.

In growing enterprises, data science departments, and research facilities data
scientists, often the only computer science or engineering experts, also have to
take care of their infrastructure. A simple solution is established fast, but
scaling is hard. A lack of expertise often results in sub-optimal or redundant
solutions or endless research.

In this work, we showed from a data scientist perspective, how to scale an existing
infrastructure with a limited budget, while, among other requirements, maintaining
extensibility. We presented our architecture and the software stack of our cluster
with job scheduling.

\bibliography{main}{}
\bibliographystyle{IEEEtran}

\end{document}